\documentclass[aps,prl,twocolumn,amsmath,amssymb,superscriptaddress,nofootinbib]{revtex4-2}

\usepackage{braket,gensymb,tikz}
\usepackage{graphicx,epstopdf}
\usepackage{soul}

\newcommand{\subcaptionOverlay}[3]{
  \begin{tikzpicture}
    \node [inner sep=0,anchor=north west]at (#3) (image) {#1};
    \draw node [black] {#2};
  \end{tikzpicture}
}

\begin{document}

\title{Raman sideband cooling of molecules in an optical tweezer array to the 3-D motional ground state}

\author{Yicheng Bao}
\author{Scarlett S. Yu}
\author{Jiaqi You}
\author{Lo\"ic Anderegg}
\affiliation{Department of Physics, Harvard University, Cambridge, MA 02138, USA}
\affiliation{Harvard-MIT Center for Ultracold Atoms, Cambridge, MA 02138, USA}

\author{Eunmi Chae}
\affiliation{Department of Physics, Korea University, Seongbuk-gu, Seoul, South Korea}

\author{Wolfgang Ketterle}
\affiliation{Harvard-MIT Center for Ultracold Atoms, Cambridge, MA 02138, USA}
\affiliation{Department of Physics, Massachusetts Institute of Technology, Cambridge, MA 02139, USA }

\author{Kang-Kuen Ni} 
\affiliation{Department of Physics, Harvard University, Cambridge, MA 02138, USA}
\affiliation{Harvard-MIT Center for Ultracold Atoms, Cambridge, MA 02138, USA}
\affiliation{Department of Chemistry and Chemical Biology, Harvard University, Cambridge, MA 02138, USA}

\author{John M. Doyle} 
\affiliation{Department of Physics, Harvard University, Cambridge, MA 02138, USA}
\affiliation{Harvard-MIT Center for Ultracold Atoms, Cambridge, MA 02138, USA}

\date{\today}

\begin{abstract}
Ultracold polar molecules are promising for quantum information processing and searches for physics beyond the Standard Model. Laser cooling to ultracold temperatures is an established technique for trapped diatomic and triatomic molecules. Further cooling of the molecules to near the motional ground state is crucial for reducing various dephasings in quantum and precision applications. In this work, we demonstrate Raman sideband cooling of CaF molecules in optical tweezers to near their motional ground state, with average motional occupation quantum numbers of $\bar{n}_{x}=0.16(12)$, $\bar{n}_{y}=0.17(17)$ (radial directions), $\bar{n}_{z}=0.22(16)$ (axial direction) and a 3-D motional ground state probability of $54\pm18\%$. This paves the way to increase molecular coherence times in optical tweezers for robust quantum computation and simulation applications.
\end{abstract}

\maketitle
\section{Introduction}
Owing to their intrinsic electric dipole moment and rich internal structures, ultracold polar molecules are an attractive platform for quantum simulation of strongly interacting many-body dynamics~\cite{micheli2006toolbox,gadway2016strongly, baranov2008theoretical, wall2015magnetism} and for quantum computation~\cite{demille2002quantum,yelin2006dipolarQC, karra2016paramagnetic,sawant2020qudits,ni2018dipolar}. In addition, the long interaction times available with trapped ultracold molecules can greatly aid searches for physics beyond the Standard Model~\cite{doyle2022ultracold,kozyryev2017precision,hutzler2020polyatomic,anderegg2023quantum,kozyryev2021enhanced,norrgard2019nuclear,hao2020nuclear}. Recent advances include loading of optical tweezer arrays~\cite{anderegg2019optical,zhang2022optical}, demonstrating long rotational coherence times of molecular qubits~\cite{burchesky2021rotational,gregory2023second,park2023extended}, and realizing rotational state spin exchange between individual molecules in both optical tweezers~\cite{holland2022demand,bao2022dipolar} and optical lattices~\cite{yan2013observation,li2023tunable,christakis2023probing}. Molecules are made ultracold typically through one of two methods, assembly of ultracold atomic constituents ~\cite{ni2008high,molony2014creation,takekoshi2014ultracold,park2015ultracold,guo2016creation,rvachov2017long,liu2018building,he2020coherently} or direct laser cooling~\cite{barry2014magneto,anderegg2017radio,truppe2017molecules,collopy20183d,vilas2022CaOHMOT}. Laser-cooled molecules provide high fidelity optical detection and straightforward loading of large tweezer arrays. So far, observed rotational qubit coherence times of laser-cooled molecules have been limited by dephasing induced by thermal motion~\cite{burchesky2021rotational}. In particular, motion along the most weakly trapped direction (axial) of the optical trap dominates the dephasing in molecule-molecule electric dipolar interactions, due to the spatially dependent variation of the interaction strength. This also degrades Bell state fidelities and two-qubit gate performance~\cite{bao2022dipolar,holland2022demand}. Further cooling would reduce these decoherence mechanisms. However, it is challenging to cool a molecule in an optical trap in the absence of a narrow-line transition, as is the case with most (but not all~\cite{wu2021high}) laser cooled molecules.

Raman sideband cooling (RSC) techniques have been used to cool ions in traps~\cite{monroe1995resolved} and neutral atoms in optical lattices and optical tweezers ~\cite{vuletic1998degenerate,hamann1998resolved,perrin1998sideband,kaufman2012cooling,thompson2013coherence}. The method typically consists of two key steps. First, a motional-sideband-resolved two-photon Raman transition is driven, reducing the motional quantum number $\ket{n}$ of the harmonic trapping potential, while also changing the internal state of the atom. Second, optical pumping restores the internal state of the atom without affecting the motional state. By repeatedly applying these two steps, the trapped atom can be driven to the motional ground state of the trapping potential. To reach the motional ground state using RSC, it is crucial to optically resolve the motional sidebands in all three dimensions and to achieve efficient optical pumping.

In this work, we realize Raman sideband cooling (RSC) of ultracold CaF molecules trapped in optical tweezers and cool the molecules to near the three-dimensional motional ground state in $100\,\text{ms}$. The experimental approach is guided by the proposal of Caldwell and Tarbutt~\cite{caldwell2020sideband}, whereby a high magnetic field decouples the electron spin and nuclear spin from the molecular rotation, significantly reducing the number of quantum states in the RSC scheme. For molecules trapped in deep optical tweezers, differential ac stark shifts can broaden the Raman transition, making it challenging to resolve the motional sidebands. With our choice of states at high magnetic field, these difficulties are overcome, resulting in a simple scheme for applying RSC on molecules. Following the application of RSC, we characterize the average motional occupation number of the molecules using Raman sideband thermometry (RST).  In addition, after RSC we observe increased coherence of Raman Rabi oscillations between motional states.

\begin{figure*}[!htbp]
\includegraphics[width=\textwidth]{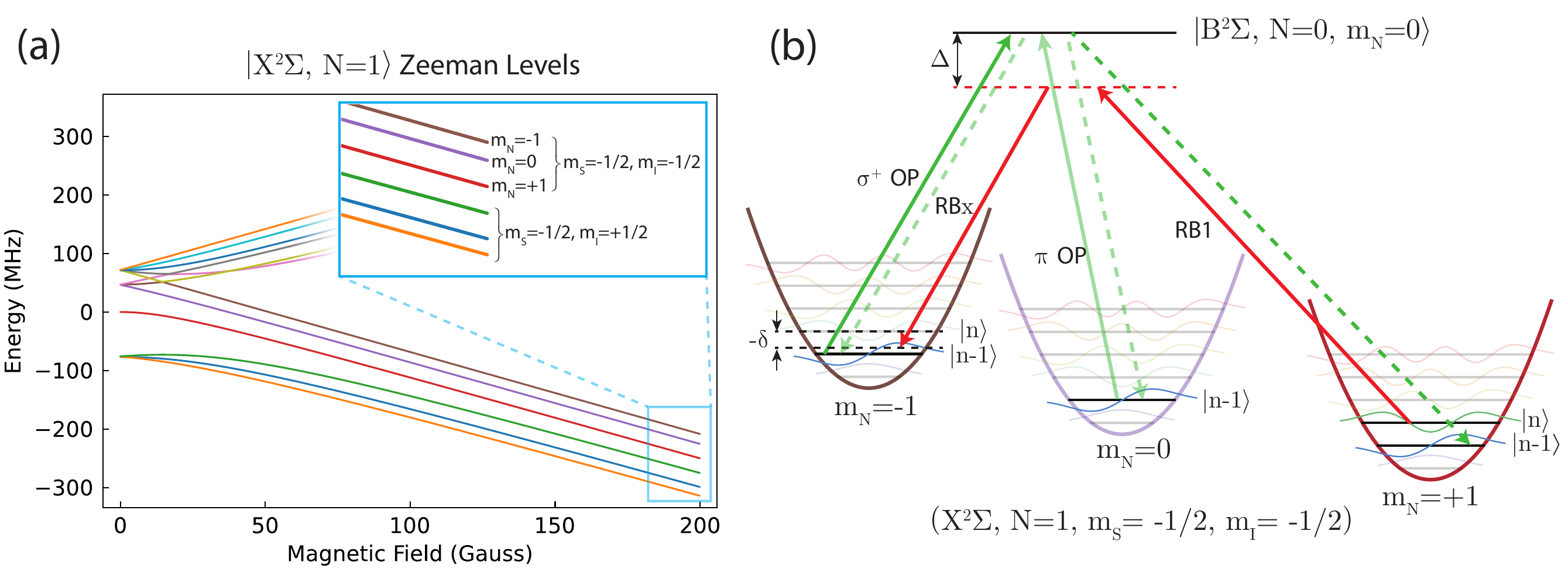}
\caption{(a) Zeeman energy levels of the $\ket{X^{2}\Sigma, N=1}$ manifold. The $(m_{S}=-1/2,m_{I}=-1/2)$ subspace used for RSC in this work is highlighted in the blue inset. (b) Motional eigenstates $\ket{n}$ in a harmonic trap. Red arrows labeled RB1 and $\text{RB}x$ ($x=2,3\text{ or }4$) represent the Raman transition driven to lower the motional quantum number $n$ by $1$ for a specific trap axis. The green arrows represent the $\sigma +$ and the $\pi$ optical pumping transitions. The dashed green arrow indicates spontaneous emission. The $\pi$-polarized transition is driven to prevent accumulation of population in $\ket{X^{2}\Sigma, N=1, m_{N}=0}$ state during OP. Note the negative sign in front of the Raman two-photon detuning $\delta$, indicating cooling sidebands have a negative $\delta$ value. }
\label{fig:1}
\end{figure*}

\section{Raman Sideband Cooling, Thermometry, and the Experimental Approach}

RSC necessarily requires many steps of optical pumping and two-photon Raman transitions. However, this is challenging with molecules due to the large number of internal states, which can lead to population leakage to states outside the cooling cycle. In order to minimize leakage, a high (${\sim}200\,\text{G}$) magnetic field is applied, resolving and isolating the state manifold for RSC~\cite{caldwell2020sideband} by decoupling the electron spin $S$ and nuclear spin $I$ from rotational angular momentum $N$ (Fig.~\ref{fig:1}~(a)). These decoupled angular momentum eigenstates of the molecule are written as $\ket{S, m_{S}}\ket{I,m_{I}}\ket{N,m_{N}}$, where $m_{S}$ ($m_{I}$) is the projection of the electron (nuclear) spin angular momentum onto the quantization axis. When driving optical pumping transitions between the $X^{2}\Sigma$ and $B^{2}\Sigma$ electronic states in CaF at this magnetic field strength, the transitions that change $m_{I}$ and $m_{S}$ are suppressed. This reduces the number of scattered photons needed for cooling and therefore results in good state-preserving efficiency and lower heating~\cite{caldwell2020sideband}. A small subspace $(m_{S}=-1/2,m_{I}=-1/2)$ in the $X^{2}\Sigma$ electronic ground state is used, with only three relevant RSC states, $\ket{m_{N}=0,\pm1}$ (Fig.~\ref{fig:1}~(b)). For CaF, the Franck-Condon factors of the transitions between the $X^{2}\Sigma$ and $B^{2}\Sigma$ states are highly diagonal (${\sim}0.998$), so that the loss into excited vibrational states is minimal.

In our application of RSC, cooling is initiated outside of the Lamb-Dicke (LD) regime due to a combination of factors. These factors are dominantly the relatively light mass of CaF, the low trap frequency in both the radial and axial directions of the tweezer trap, and the initial temperature set by the limits of $\Lambda$-enhanced gray molasses cooling (${\sim}60\,\mu\text{K}$ in a ${\sim}k_{B}\times1.3\,\text{mK}$ deep trap, equivalent to $\bar{n}_{\text{radial}}>6$ and $\bar{n}_{\text{axial}}>40$). We perform the cooling by setting the two-photon detuning $\delta$ of the Raman lasers to address higher-order motional sidebands ($|\Delta n|>1$, where $\Delta n$ is the change of the motional quantum number) in the initial cooling steps~\cite{yu2018motional}. This allows the removal of more motional energy in a single RSC cycle while leaving the amount of heating from the optical pumping step unchanged. We then switch to driving only one or two states to finish the cooling process.

We employ RST to determine the temperature of the molecules (or, equivalently, the average motional occupation number $\bar{n}$). In this technique, Raman transitions (the same that are used in RSC) are driven for a fixed duration, and the population transferred to $\ket{m_{N}=-1}$ is then measured using a state-sensitive detection scheme. The two photon detuning $\delta$ (Fig.~\ref{fig:1}~(b)) of the Raman transition is scanned, forming a Raman sideband thermometry spectrum. Generally, an RST spectrum contains a center peak at the carrier frequency ($\Delta n = 0$), with ``cooling'' ($\Delta n<0$) and ``heating'' sideband peaks ($\Delta n>0$) at lower and higher frequencies. The average occupation number of the motional states, $\bar{n}$, can be related to the ratio of the peak heights of the first cooling sideband to the first heating sideband, $\alpha$, using the relation $\bar{n}=\frac{\alpha}{1-\alpha}$~\cite{monroe1995resolved,yu2018motional}. The signature of a low temperature sample is the strong asymmetry between the peak height of the cooling and heating sidebands.

\begin{figure*}[!htbp]
\includegraphics[width=0.6\textwidth]{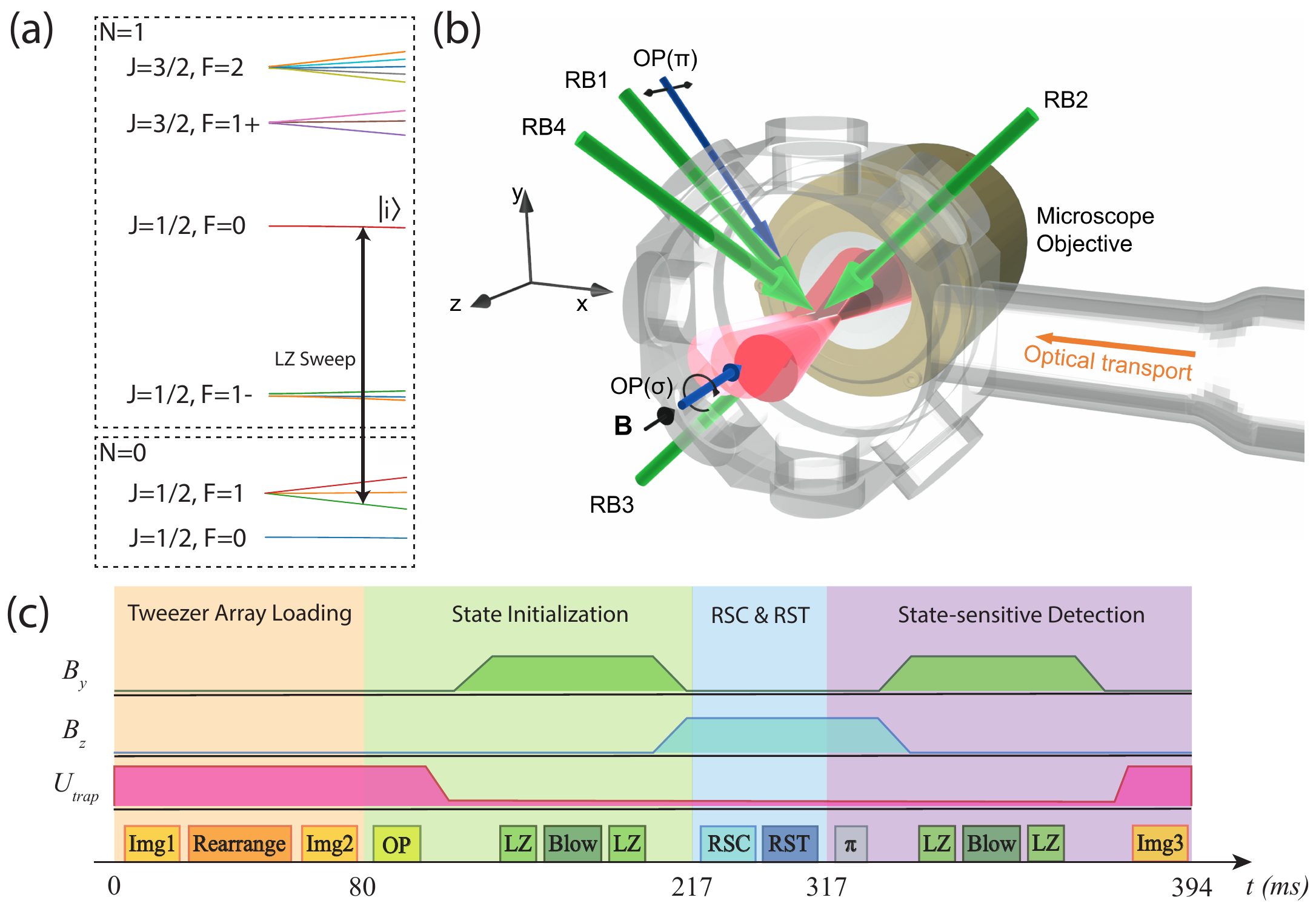}
\caption{(a) Energy levels of the first two rotational manifolds of the CaF $\ket{X^{2}\Sigma}$ electronic ground state at low magnetic fields (not to scale). LZ sweeps are used to transfer population between the $\ket{i}\equiv\ket{N=1,F=0,m_{F}=0}$ state and the $\ket{N=0,F=1,m_{F}=-1}$ state for state preparation and detection. (b) Configuration of Raman beams and optical pumping beams used in RSC. RB1 and RB2 address the radial motion in $x$-direction, RB1 and RB3 address the radial motion in $y$-direction, and RB1 and RB4 mainly address the axial motion in $z$-direction. The RB1 path is also be employed to drive a motion-insensitive Raman $\pi$ pulse for population swapping between $\ket{m_{N}=\pm1}$ states. (c) Experimental sequence of tweezer array preparation, state initialization, RSC, RST, and state-sensitive detection. $B_{y}$ and $B_{z}$ are the applied magnetic field along $y$ direction and $z$ direction. $U_{trap}$ is the trap depth of the optical tweezer trap. $\pi$ denotes the motion-insensitive Raman $\pi$-pulse used to transfer the population from $\ket{m_{N}=-1}$ to $\ket{m_{N}=+1}$.}
\label{fig:2}
\end{figure*}

Our experimental approach can be described as having four phases: loading of molecules into a tweezer array, state initialization, RSC, and RST.

CaF molecules are loaded into an optical tweezer array using the method described in previous work~\cite{bao2022dipolar}. In brief, a buffer gas beam source loads a radio-frequency magneto-optical trap (RF MOT) with CaF molecules~\cite{anderegg2017radio}. The molecules are transferred into a 1-D optical lattice, which transports them into a glass cell~\cite{bao2022fast}, where they are loaded into individual optical tweezers in a 1-D array using $\Lambda$-enhanced gray molasses cooling. A compact array of molecules is achieved by first $\Lambda$-imaging~\cite{cheuk2018lambda} each tweezer trap (Img 1 in Fig.~\ref{fig:2}~(c)), and, using this detection result, rearranging the loaded tweezers in real time towards one side of the array. A subsequent $\Lambda$-imaging pulse (Img 2 in Fig.~\ref{fig:2}~(c)) is then used to confirm the creation of the rearranged array.

The second phase is state initialization. After $\Lambda$-imaging, the molecular population is distributed among multiple hyperfine states of the $X^{2}\Sigma, N=1$ rotational manifold (Fig.~\ref{fig:2}~(a)). In order to initialize the molecules in the required single quantum state for RSC, we first optically pump them into the $\ket{i}\equiv\ket{N=1, F=0, m_{F}=0}$ state. Next, the trap depths of the tweezers are ramped down from ${\sim}k_{B}\times1.3\,\text{mK}$ to ${\sim}k_{B}\times220\,\mu\text{K}$. This adiabatically cools the molecules from a temperature of ${\sim}60\,\mu\text{K}$ to ${\sim}25\,\mu\text{K}$. A small bias magnetic field along the $y$-axis, $B_{y}{\sim}2.8\,\text{G}$, is then applied to lift the degeneracy between $m_{F}$ states. This allows the use of a microwave Landau-Zener (LZ) sweep to shelve the population into the $\ket{N=0,F=1, m_{F}=-1}$ state. A laser pulse resonant with the $\ket{X^{2}\Sigma,N=1}-\ket{A^2\Pi_{1/2},J=1/2}$ transition removes the residual molecules left in the $N=1$ manifold while not affecting the shelved population in the $N=0$ manifold. After this, we apply a LZ sweep to transfer the molecules back into the $\ket{i}$ state. In the final step of state initialization, a magnetic field of $B_{z}{\sim}200\text{G}$ is applied, and $B_{y}$ is reduced to zero. These fields are ramped so as to adiabatically rotate the quantization axis and prepare the molecules in the initial state for Raman sideband cooling, $\ket{m_{N}=+1}\equiv\ket{m_{I}=-1/2, m_{S}=-1/2, m_{N}=+1}$. Starting from here, RSC and RST can be applied.

The third phase is RSC. The RSC mechanism fundamentally relies on the momentum kick imparted in the Raman transition step, which lies in the direction defined by the difference of the $\mathbf{k}$-vectors between two Raman beams. To make a Raman transition couple to motion in all three directions of the harmonic potential, we use three pairs of laser beams separated by specific angles (Fig.~\ref{fig:2}~(b)). Four beams are employed (RB1 \& RB$x$, $x=\{2,3,4\}$). The RB1 beam is the shared arm for all three pairs of RSC beams, i.e, RB$2-4$ are used in conjunction with RB1 to address the three orthogonal axes of the trap. All beams are set to have a single photon detuning of $\Delta{\sim}2\pi\times44\,\text{GHz}$, red-detuned from the $\ket{X^{2}\Sigma,N=1}$ to $\ket{B^{2}\Sigma,N=0}$ transition (Fig.~\ref{fig:1}~(b)), and each is separately prepared with an acousto-optic modulator to define both its frequency and intensity. The two arms of each pair of beams are set with a frequency difference ${\sim}41\,\text{MHz}$ (defined as the carrier frequency), chosen to match the energy difference between the RSC ground states, $\ket{m_{N}=\pm 1}$, at $B_{z}{\sim}200\,\text{G}$. The RB$2-4$ beams in our experiment are turned on as a square pulse, whereas RB1 is a Blackman pulse, which greatly reduces off-resonant driving and helps to resolve closely spaced sidebands. 

In the OP steps of RSC, both $\sigma^+$ and $\pi$ transitions are driven to create a dark state in $\ket{m_{N}=+1}$. Two OP beams are used, where one propagates perpendicular to the magnetic field direction (linearly polarized along the magnetic field so as to address the $\pi$ transition), and the other propagates along the magnetic field direction (circularly polarized to address the $\sigma^+$ transition) (Fig.~\ref{fig:2}~(b)). During each OP step, ${\sim}3$ photons on average are estimated to be scattered based on calculated branching ratios.

The fourth and final phase of the experiment is RST. (RST is also used to measure the average initial temperature of loaded molecules before RSC, as will be explained in the next section of this paper.)  To perform RST, we scan the two-photon detuning $\delta$ of the Raman beams in order to record a spectrum of motional sidebands. To read out the result of RST (population transferred to $\ket{m_{N}=-1}$), a state-sensitive detection method is employed. The first step is to drive a $\pi$-pulse on the motion-insensitive Raman transition to transfer the population from $\ket{m_{N}=-1}$ to $\ket{m_{N}=+1}$ (see supplemental materials). Subsequent detection of population in $\ket{m_{N}=+1}$ is done in a similar way to the state initialization procedure. Specifically, we ramp the magnetic fields back to $B_{z}=0$ and $B_{y}{\sim}2.8\,\text{G}$, apply a LZ sweep to transfer population from state $\ket{i}$ into the $N=0$ manifold, remove the remaining population in the $N=1$ manifold, and finally apply a LZ sweep to transfer the shelved $N=0$ molecules back to $\ket{i}$. The tweezers are then ramped back to full depth and a third $\Lambda$-imaging pulse (Img 3 in Fig.~\ref{fig:2}~(c)) is used to detect the molecules. The fraction of molecules in $\ket{m_{N}=-1}$ is calculated by dividing the number of detected molecules in Img 3 by the number detected in Img 2.

\section{Experimental Results}

\begin{figure*}[!htbp]
\begin{minipage}{0.32\textwidth}
\subcaptionOverlay{\includegraphics[width=1\textwidth]{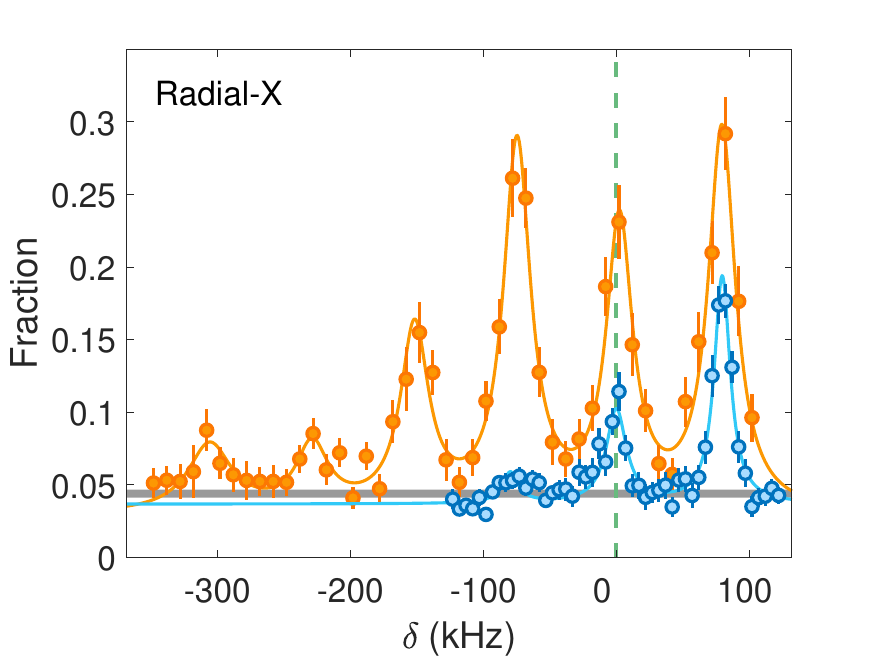}}{(a)}{-1ex,0ex}
\end{minipage}
\begin{minipage}{0.32\textwidth}
\subcaptionOverlay{\includegraphics[width=1\textwidth]{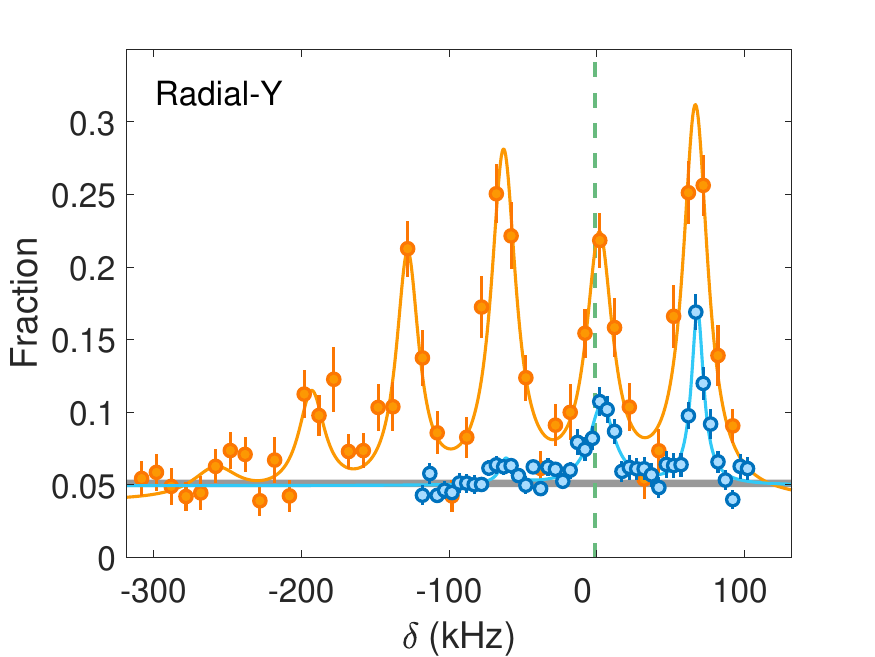}}{(b)}{-1ex,0ex}
\end{minipage}
\begin{minipage}{0.32\textwidth}
\subcaptionOverlay{\includegraphics[width=1\textwidth]{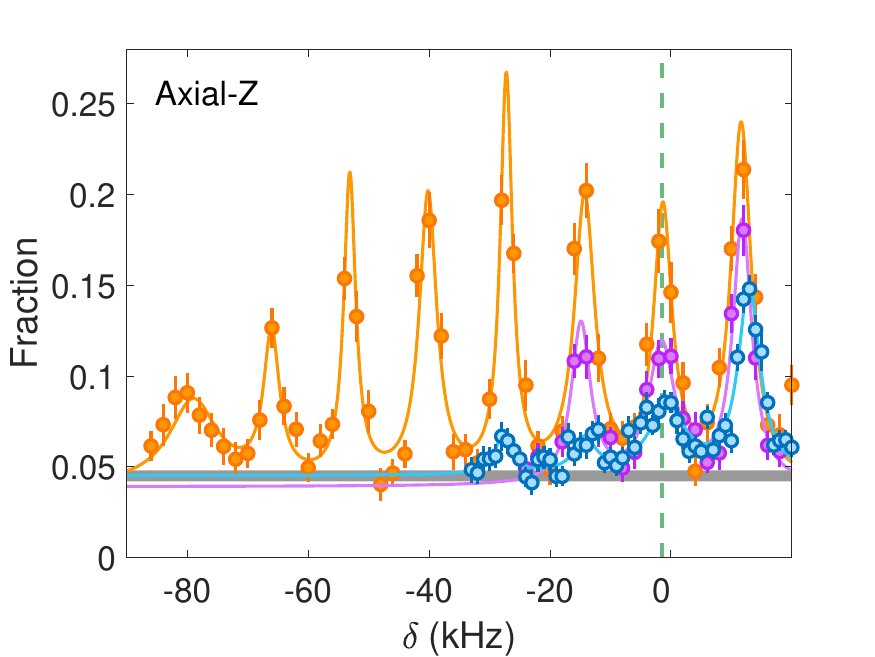}}{(c)}{-1ex,0ex}
\end{minipage}

\caption{Raman sideband thermometry spectra: (a) radial $x$ direction; (b) radial $y$ direction; and (c) axial $z$ direction. Blue curves are the spectra obtained after RSC, orange curves are the spectra obtained before RSC. For the axial $z$ direction, an additional purple curve represents the spectrum obtained without the last step of RSC applied (partial cooling). Green dashed line indicates the carrier frequency for each spectrum. Gray bands indicate the baseline for the spectra measured at the far detuned limit.}
\label{fig:3}
\end{figure*}

To characterize the initial temperature of molecules loaded into tweezers, before RSC, we create multiple realizations of tweezer loading and state initialization, Just after state initialization, RST is applied and an average initial temperature is determined. In the RST data (before RSC), motional sidebands are clearly resolved in all three directions (see Fig.~\ref{fig:3}, orange data). The data show cooling sidebands (to the left of the carrier peak) of up to $\Delta n=-4$ in the radial directions and $\Delta n=-6$ in the axial direction, where $\Delta n$ is the difference in the motional quantum number between the carrier and the sideband. These values of $|\Delta n|>1$ indicate that the cooling must start outside of LD regime. The resolved sidebands are used to determine the trap frequencies for creating the RSC recipe, in our case, $\omega_{x}=2\pi\times75(1)\,\text{kHz}$, $\omega_{y}=2\pi\times65(2)\,\text{kHz}$ (radial directions) and $\omega_{z}=2\pi\times13.6(3)\,\text{kHz}$ (axial directions).

Starting again with molecules in this initial loading and state initialization condition, we apply the RSC sequence. As we start outside of the LD regime, we initially address $\Delta n=-3$ motional sidebands for radial cooling and $\Delta n=-5$ motional sidebands for axial cooling. As cooling proceeds we switch over to driving only the first and second order motional sidebands to finish the cooling. The cooling sequence is detailed in the supplementary material. After RSC, RST is performed. The data show that cooling sidebands are strongly suppressed relative to the heating sidebands in all three directions (Fig.~\ref{fig:3}, blue curves). The average motional occupation numbers are $\bar{n}_{x}=0.16(12)$, $\bar{n}_{y}=0.17(17)$ (radial directions) and $\bar{n}_{z}=0.22(16)$ (axial direction), giving a probability for preparing the molecules in the motional ground state of $54\pm18\%$.

\begin{figure*}[!htbp]
\begin{minipage}{0.32\textwidth}
\subcaptionOverlay{\includegraphics[width=1\textwidth]{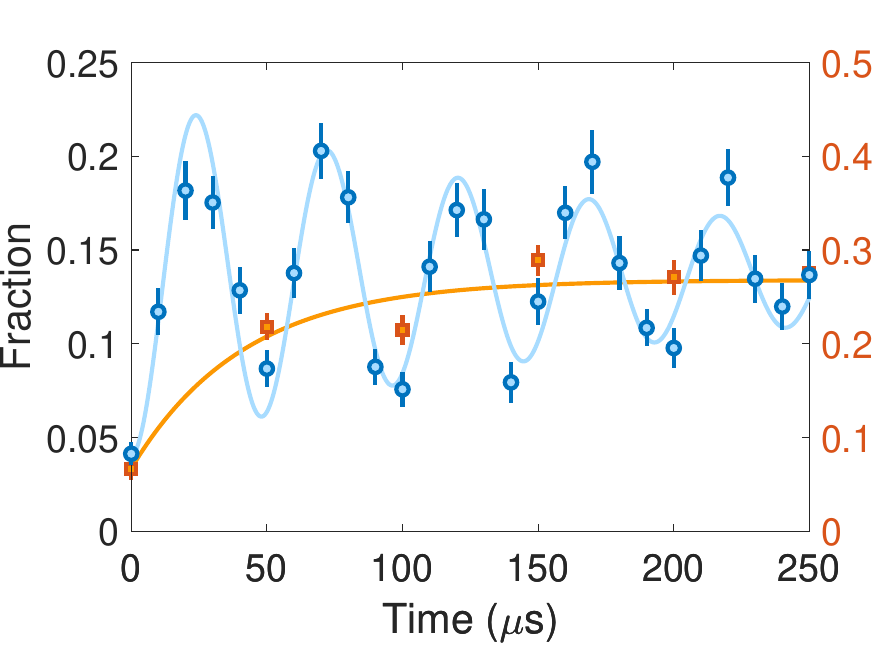}}{(a)}{-1ex,0ex}
\subcaptionOverlay{\includegraphics[width=1\textwidth]{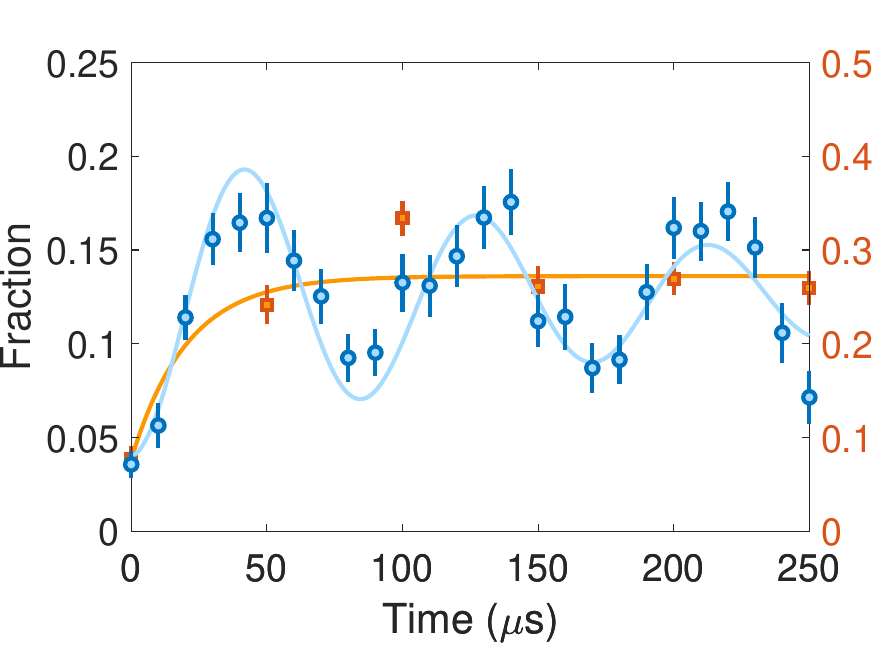}}{(d)}{-1ex,0ex}
\end{minipage}
\begin{minipage}{0.32\textwidth}
\subcaptionOverlay{\includegraphics[width=1\textwidth]{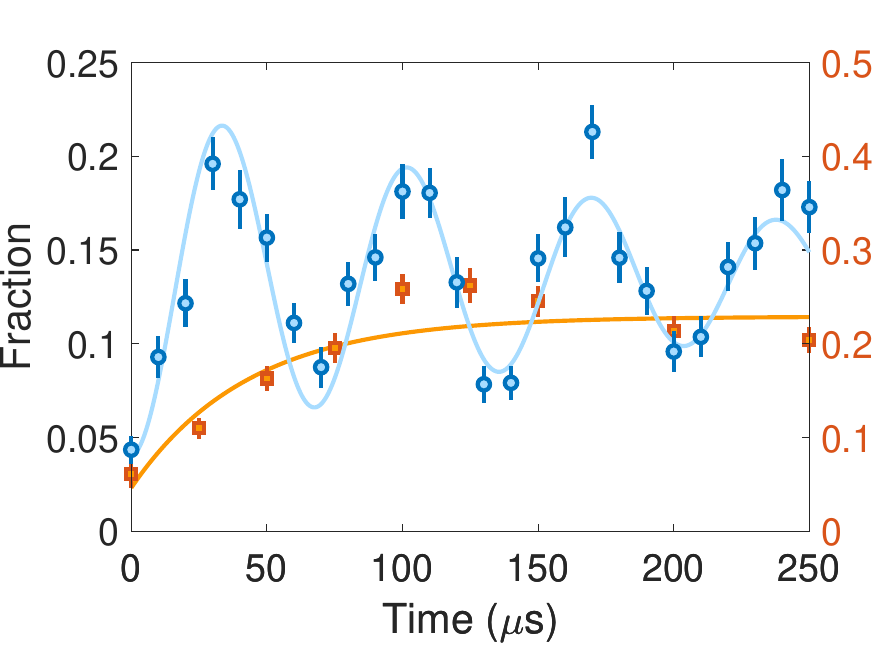}}{(b)}{-1ex,0ex}
\subcaptionOverlay{\includegraphics[width=1\textwidth]{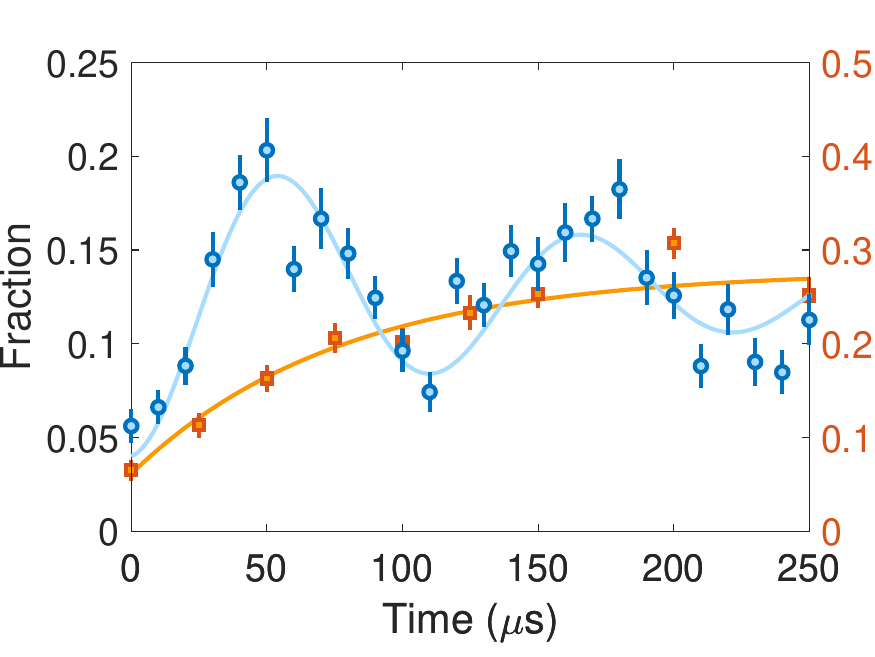}}{(e)}{-1ex,0ex}
\end{minipage}
\begin{minipage}{0.32\textwidth}
\subcaptionOverlay{\includegraphics[width=1\textwidth]{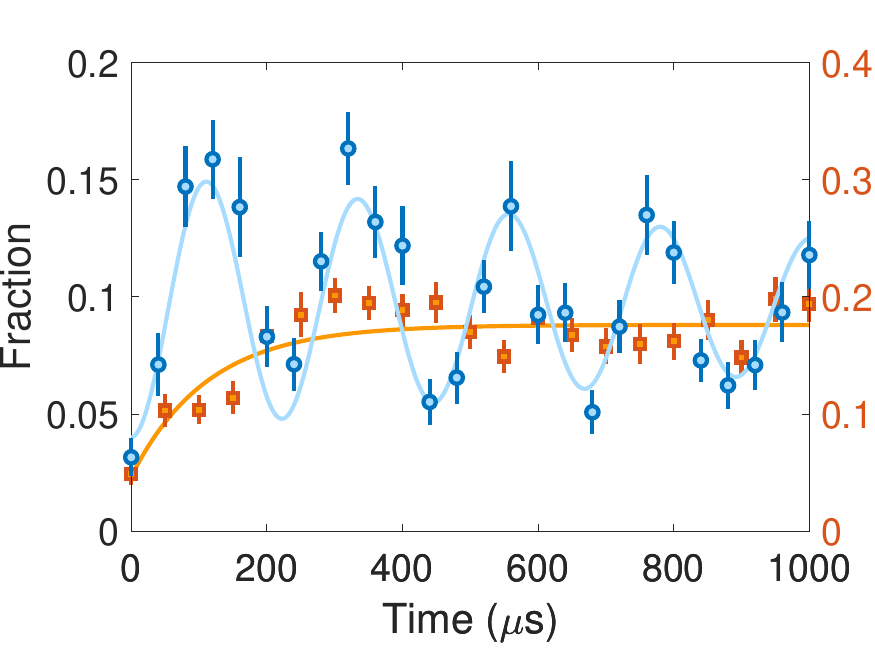}}{(c)}{-1ex,0ex}
\subcaptionOverlay{\includegraphics[width=1\textwidth]{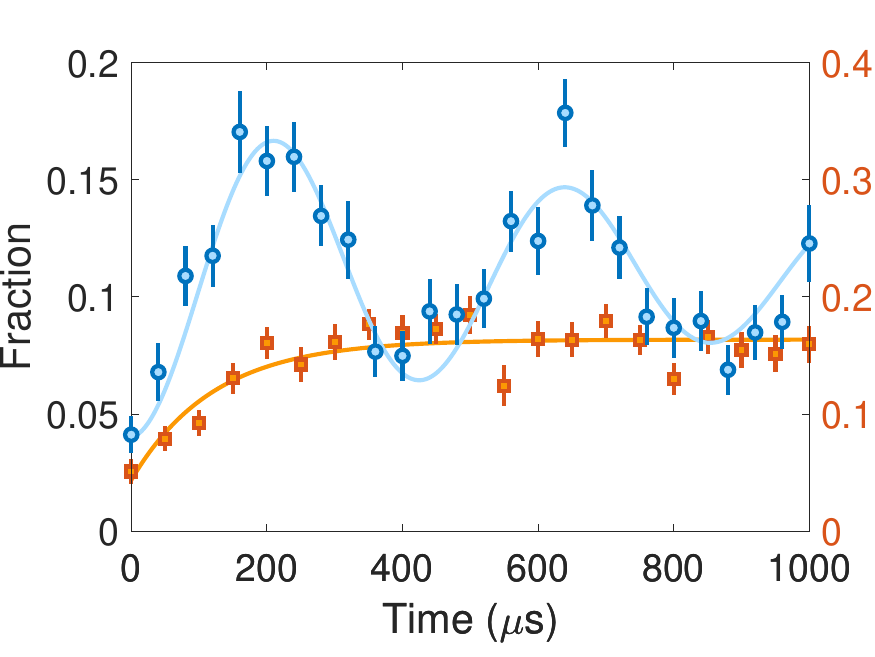}}{(f)}{-1ex,0ex}
\end{minipage}

\caption{Rabi oscillations of the carrier transition: (a) radial $x$ direction; (b) radial $y$ direction; and (c) axial $z$ direction. Rabi oscillations of the first heating sideband: (d) radial $x$ direction; (e) radial $y$ direction; and (f) axial $z$ direction. In each figure, the data showing rapid dephasing before the application of RSC are plotted as orange squares, fitted to an exponential decaying model. Coherent Rabi oscillations after the application of RSC are plotted as blue circles, fitted to a sine model with exponentially decaying amplitude.}
\label{fig:4}
\end{figure*}

The coherence of Rabi oscillations driven on the sideband or carrier transitions can also provide information on the temperature of molecules. Since the Rabi oscillation frequencies for transitions with $\Delta n=0,\pm1$ depend on $n$, if the molecules occupy a large number of motional $\ket{n}$ states, rapid dephasing would be expected. We observe this behavior when attempting to drive Rabi oscillations in the initial configuration (without applying RSC, Fig.~\ref{fig:4}, orange data). In contrast, if the molecules occupy a single $\ket{n}$ state, coherent Rabi oscillations should be observable. After applying RSC, coherent Raman Rabi oscillations on both the carrier and the first heating sideband are clearly observed in all three directions  (see Fig.~\ref{fig:4}, blue data), showing this effect of the cooling.

In order to study the efficiency of our RSC procedure, we varied the number of Raman and OP steps. As more RSC steps are applied, apparent increased loss of molecules is observed. The survival rate is around $40\%$ after the application of the full RSC procedure. In order to understand the cause of this observation, we turn off the state-sensitivity of the detection in order to detect all the molecules in the $N=1$ manifold. We verify that the total population in the $N=1$ manifold remains the same within the measurement error, indicating that this loss is not caused by loss of molecules from the trap, but by decay into other $(m_{S},m_{I})$ subspaces. We attribute this to residual mixing between different $(m_{S},m_{I})$ subspaces~\cite{caldwell2020sideband}. The observed loss scales with the number of photons scattered during the optical pumping steps of RSC, and a large number of RSC cycles are applied due to the high initial temperature of the molecules. This loss could be mitigated in the future by applying RSC simultaneously on four $m_{S}$ and $m_{I}$ subsets of states~\cite{caldwell2020sideband}. In an alternative method, one may be able to ramp down the magnetic field in the middle of the RSC pulse sequence and apply $m_{F}$ optical pumping to re-initialize the molecules in $\ket{N=1,F=0,m_{F}=0}$ state. Since the molecules are already partially Raman sideband cooled, a smaller number of additional RSC cycles would be needed in the final cooling step.

\section{Conclusion}

In summary, we have demonstrated Raman sideband cooling of CaF molecules in an optical tweezer array. Motional sidebands in all three orthogonal directions are spectrally resolved. After applying Raman sideband cooling, resolved sideband thermometry measures the temperature of the molecules in the tweezers, indicating a 3-D motional ground state probability of $54\pm18\%$. Dephasing of Raman Rabi oscillations is highly suppressed after cooling. Currently, the cooling performance is mainly limited by the off-resonance driving in the Raman steps and heating from the optical pumping steps. To overcome these limitations, the high magnetic field could be better stabilized to allow for lower Raman Rabi frequency and reduced off-resonance driving. In addition, by choosing the high magnetic field direction perpendicular to the trap axial axis, the heating effect from the $\sigma^+$ optical pumping could be suppressed~\cite{caldwell2020sideband}. The successful application of Raman sideband cooling on laser-coolable molecules in  optical tweezers is an important step toward increasing molecular rotational coherence times and reducing motional dephasing for two qubit dipolar gate operations. This work increases the possibilities and power of molecular optical tweezer arrays for use in quantum computation quantum simulation applications and precision searches for physics beyond the Standard Model. We expect that this cooling approach can also be applied to laser-coolable polyatomic molecules in optical tweezers or lattices~\cite{hallas2022optical}.

We note other work on RSC of CaF molecules in optical tweezers~\cite{lu2023raman}.

\section{Acknowledgments}
\begin{acknowledgments}
This material is based upon work supported by the U.S. Department of Energy, Office of Science, National Quantum Information Science Research Centers, Quantum Systems Accelerator. Additional support is acknowledged from AFOSR, AOARD and ARO. SY acknowledges support from the NSF GRFP. LA and SY acknowledge support from the HQI. EC acknowledges support from the NRF of Korea (2021M3H3A1085299, 2022M3E4A1077340, 2022M3C1C8097622). We thank Nathaniel B. Vilas for reading and editing of the manuscript.
\end{acknowledgments}

\clearpage
\section{Supplemental Material}

\subsection{Raman beam alignment}
To align the Raman beam paths onto the molecules, we drive a motion-insensitive Raman transition using two frequency components that are combined into a single polarization-maintaining fiber and delivered to the molecules through one of the RB$1-4$ beam paths. By scanning the relative frequency between the two frequency components, maximal population transfer between the $\ket{m_{N}=\pm1}$ is observed at the carrier resonance. We then maximize the Rabi frequency of this motion-insensitive Raman transition by tuning the alignment of the Raman beam path onto the molecules, ensuring uniform illumination across the array. A subset of six tweezers in the full array is used in this work. Over this subset, we measure a root-mean-squared (RMS) deviation of Rabi frequency of less than $3\%$, verifying the uniform illumination over the tweezer subset. This procedure is repeated for each of the RB$1-4$ beam paths to finish the alignment. 

We also use this motion-insensitive Raman transition as a convenient way to transfer population between $\ket{m_{N}=\pm1}$. This is utilized in the experiment sequence to realize state-sensitive detection on $\ket{m_{N}=-1}$ state required by RST.

\subsection{Avoided crossing in magnetic field ramp}

\begin{figure}[!htbp]
\includegraphics[width=\columnwidth]{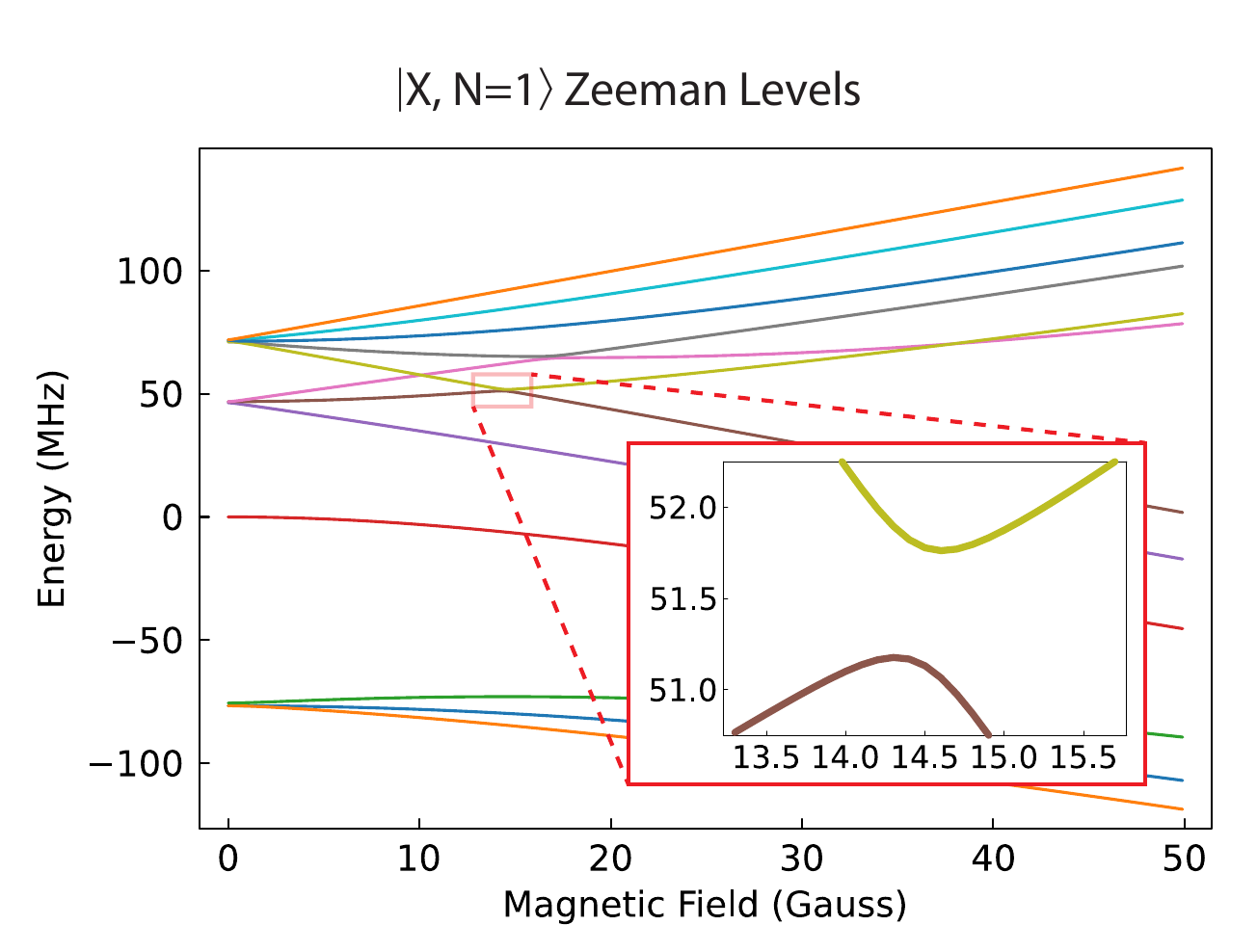}
\caption{Avoided crossing between $\ket{F=2,m_{F}=-2}$ state and $\ket{F=1+, m_{F}=0}$ state around magnetic field $B_{\text{cross}}\approx14.5\,\text{G}$.}
\label{fig:S1}
\end{figure}

To prepare the molecules in the $(m_{S}=-1/2, m_{I}=-1/2)$ subspace for RSC, it is less ideal to start from $\ket{F=2,m_{F}=-2}$ state and ramp the magnetic field to prepare $\ket{m_{I}=-1/2,m_{S}=-1/2, m_{N}=-1}$, since the ac stark shift from the tweezer trap can cause mixing between $\ket{F=2,m_{F}=-2}$ and $\ket{F=1+,m_{F}=0}$ at finite magnetic field, creating an avoided crossing when magnetic field is around $15\,\text{G}$ (see Fig.~\ref{fig:S1} inset). Due to technical limitation, our magnetic field ramp rate has a similar time scale as the energy splitting of the avoided crossing, causing leakage into $\ket{F=1+,m_{F}=0}$ states.

\subsection{Raman sideband cooling recipe}
\begin{figure}[!htbp]
\includegraphics[width=\columnwidth]{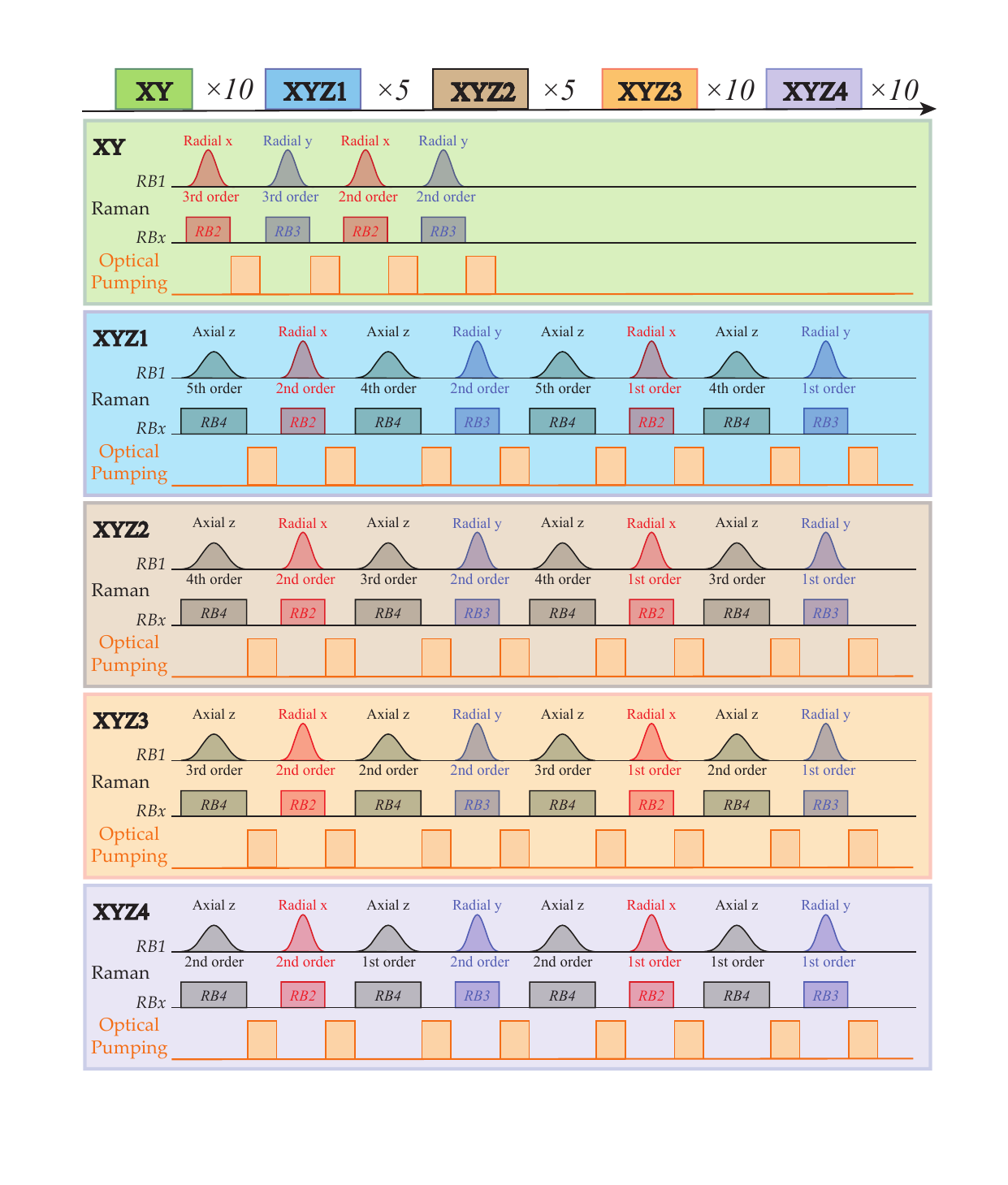}
\caption{Raman sideband cooling pulse sequences used in this work.}
\label{fig:S2}
\end{figure}

The Raman sideband cooling pulse sequence used in this work is shown in Fig.~\ref{fig:S2}. The pulse shape of RB1 is Blackman while those of RB$2-4$ are rectangular. Optical pumping (OP) pulses in the figure apply to both $\pi$ OP and $\sigma+$ OP beams.

\subsection{Raman sideband thermometry}
The Rabi frequency of a Raman transition between motional states $\ket{n}$ and $\ket{n'}$ is denoted as $\Omega_{R}^{n,n'}$. It is related to the Rabi frequency of the carrier transition by a factor denoted as $M_{n,n'}$.
\begin{equation}
\begin{split}
\Omega_{R}^{n,n'}=M_{n,n'}\Omega_{R}^{0}\\
\end{split}
\end{equation}
An analytical form of $M_{n,n'}$ is given by~\cite{wineland1998experimental}
\begin{equation}
\begin{split}
M_{n,n'}&=\braket{n|e^{i\eta^{R}(\hat{a}+\hat{a}^\dagger)}|n'}\\
&=e^{-(\eta^{R})^2/2}\sqrt{\frac{n_{<}!}{n_{>}!}}(\eta^{R})^{|n-n'|}L_{n_{<}}^{|n-n'|}((\eta^{R})^2)\\
\end{split}
\end{equation}
Where $\eta^{R}$ is the Lamb-Dicke parameter of the Raman transition. It is defined as $\eta^{R}\equiv \Delta k x_{0}$, where $\Delta k$ is the wave-vector difference between the two Raman beams and $x_{0}=\sqrt{h/(2m\omega)}$ is the spread of the ground-state wave-function of the molecule. In our experiment, we have $\eta_{x}^{R}=0.57,\eta_{x}^{R}=0.61,\eta_{z}^{R}=0.62$. $n_{<(>)}$ is the smaller(larger) one between $n$ and $n'$. $L^{\alpha}_{n}(x)$ is the generalized Laguerre polynomial.

The population transferred after a $\Delta n=\pm1$ sideband thermometry pulse is $P_{\pm}$.
\begin{equation}
\begin{split}
P_{\pm}&=\sum_{n=0}^{\infty}{p_{n}\sin^2(\Omega_{R}^{n,n\pm1} t/2)}\\
\end{split}
\end{equation}
Where $t$ is the width of the pulse applied and $p_{n}$ is the probability to find the molecule in $\ket{n}$.
Note that
\begin{equation}
M_{n,n'}=M_{n',n}
\end{equation}
and for a thermal distribution
\begin{equation}
p_{n}=(1-e^{-\frac{\hbar\omega}{k_{B}T}})e^{-\frac{n\hbar\omega}{k_{B}T}}
\end{equation}
\begin{equation}
\begin{split}
P_{+}&=\sum_{n=0}^{\infty}{p_{n}\sin^2(\Omega_{R}^{n,n+1} t/2)}\\
&=\sum_{n=0}^{\infty}{(1-e^{-\frac{\hbar\omega}{k_{B}T}})e^{-\frac{n\hbar\omega}{k_{B}T}}\sin^2(\Omega_{R}^{n,n+1} t/2)}\\
&=\sum_{m=1}^{\infty}{(1-e^{-\frac{\hbar\omega}{k_{B}T}})e^{-\frac{(m-1)\hbar\omega}{k_{B}T}}\sin^2(\Omega_{R}^{m-1,m} t/2)}\\
&=e^{\frac{\hbar\omega}{k_{B}T}}\sum_{m=1}^{\infty}{(1-e^{-\frac{\hbar\omega}{k_{B}T}})e^{-\frac{m\hbar\omega}{k_{B}T}}\sin^2(\Omega_{R}^{m-1,m} t/2)}\\
&=e^{\frac{\hbar\omega}{k_{B}T}}\sum_{n=0}^{\infty}{p_{n}\sin^2(\Omega_{R}^{n,n+1} t/2)}\\
&=e^{\frac{\hbar\omega}{k_{B}T}}P_{-}\\
\end{split}
\end{equation}
$e^{-\frac{\hbar\omega}{k_{B}T}}$ is directly linked to the sideband thermometry peak height ratio $\alpha$ and is independent of the specific choice of sideband thermometry pulse width $t$.
\begin{equation}
\begin{split}
e^{-\frac{\hbar\omega}{k_{B}T}}=\frac{P_{-}}{P_{+}}=\alpha
\end{split}
\end{equation}
The average harmonic potential occupancy number $\bar{n}$ can be determined from $\alpha$
\begin{equation}
\begin{split}
\bar{n}&=\sum_{n=0}^{\infty}{p_{n}n}\\
&=\sum_{n=0}^{\infty}{(1-e^{-\frac{\hbar\omega}{k_{B}T}})e^{-\frac{n\hbar\omega}{k_{B}T}}n}\\
&=\sum_{n=0}^{\infty}{(1-\alpha)\alpha^{n}n}\\
&=\frac{\alpha}{1-\alpha}
\end{split}
\end{equation}
Experimentally, it is desirable to choose a sideband thermometry pulse width $t$ that results in the largest $P_{+}$ and $P_{-}$, so as to increase the sensitivity of the temperature measurement.

\subsection{State initialization and detection fidelity}
We observe a combined state initialization and state-sensitive detection fidelity of $55\%$. The finite loss is attributed to the finite efficiencies of optical pumping, microwave transfer, and imaging. Potential future improvements include a $m_{F}$ optical pumping scheme where the dark state of the optical pumping is further protected by polarization selection rules~\cite{holland2022demand}.


%

\end{document}